# Non-existence of radiation damping of gravitational motions

A. LOINGER

Dipartimento di Fisica, Università di Milano

Via Celoria, 16 – 20133 Milano, Italy

**Summary.** – A rigorous, non-perturbative proof that there is no radiation damping of gravitational motions.

**1**. – It is commonly believed that any accelerated point mass sends forth gravity waves [1], and that the emission of these waves generates a *damping force* on the particle [2].

I shall now demonstrate that any gravitational motion of any point mass does not give origin to any damping term. (But without damping …).

The proof is straightforward. As is well known [3], the so-called geodesic principle is an analytical consequence of vanishing of the covariant divergence $T^{jk}{}_{;j}$ of mass tensor $T^{jk}$, together with the assumption that the mass tensor represents a continuous, incoherent "cloud of dust", i.e. that

$$(1.1) \qquad T^{jk} = \rho u^j u^k, \quad (j, k = 0, 1, 2, 3), \quad (c=1),$$

where $\rho$ (Fock's $\rho^*$) is the invariant mass density and $u^j$ is the four-velocity $dx^j/ds$. Passing to the limit of a *concentrated* mass ([3], [4]), one deduces that

$$(1.2) \qquad \frac{du^j}{ds} + \Gamma^j{}_{lk} u^l u^k = 0 \quad ,$$

i.e. the geodesic equations of a particle. The Γ's represent the gravity "proper force" created by the gravitational field of the mass point. Of course, we can add to it an external "gravity force".

Now, any geodesic line, solution of eqs. (1.2), can be written in the form

(1.3) $$y^j = a^j s + b^j ,$$

where the *y*'s are Riemann-Fermi co-ordinates, and $a^j, b^j$ are integration constants. As we see, *no damping force has influenced the motion.*

The supposed emission mechanism of gravity radiation is a trigger which fails.

(A *Minkowskian* analogue: from the conservation equations

(1.4) $$\frac{\partial(\rho u^j u^k)}{\partial x^j} = 0$$

it follows the law of rectilinear and uniform motion of a "dust" particle.)

**2**. – The above considerations can be extended quite generally to *any* mass tensor. The equations

(2.1) $$T^{jk}{}_{;j} = 0$$

give not only the differential covariant form of the energy-momentum conservation of matter, but also its *equations of motion*. The reason of this property is very simple: in general relativity the motion of matter is a "natural", "free" motion in a curved space-time, *created by the matter itself*. The plain case of sect.**1** is a good exemplification. In other terms, equations (2.1) represent, as it were, a sort of general geodesic principle. Consequently, *the motion of matter cannot generate any radiation damping*. An electromagnetic analogy is given by the *rectilinear and uniform* motions of charge distributions *in a Minkowski space-time*.

In 1981 Papapetrou and Linet [5] considered a bounded distribution of perfect fluid in *slow* motion ($v \ll c$), under the assumption that the gravity field is *weak* everywhere. They solved the problem approximately, with a perturbative method, by choosing Minkowski space-time as background world. These authors used eqs. (2.1) to determine, at each stage of perturbative expansion, the radiation reaction force on the fluid. They found a non-zero value for this force after the post-post-Newtonian approximation. I affirm that their result is deceptive, it is due entirely to the *perturbative* treatment, exactly as in Scheidegger's treatment [2]. (The approximation methods which start from a given, "rigid" metric are not always reliable because they violate a fundamental characteristic of general relativity [6]).

Our conclusion could be foreseen: the hypothetical gravity waves do not possess energy and momentum, and therefore their supposed emission cannot generate any reaction force [7].

The Michelson interferometer gave the decisive experimental confirmation of the non-existence of the cosmic ether, it will give the decisive experimental confirmation of the non-existence of the gravitational radiation [8].

> "Lang ist die Zeit, es ereignet sich aber das Wahre."
> From Hölderlin's *Mnemosyne*